\documentclass[conference]{IEEEtran}

\usepackage{cite}
\usepackage{amsmath,amssymb,amsfonts}
\usepackage{algorithmic}
\usepackage{graphicx}
\usepackage{textcomp}
\usepackage{xcolor}
\def\BibTeX{{\rm B\kern-.05em{\sc i\kern-.025em b}\kern-.08em
    T\kern-.1667em\lower.7ex\hbox{E}\kern-.125emX}}
\usepackage{flushend} 
\usepackage{hyperref} 
\usepackage{orcidlink} 

\newcommand{\midsize}{\fontsize{8}{10}\selectfont}
\usepackage{courier} 
\usepackage{listings, xcolor}

\definecolor{darkblue}{rgb}{0.0,0.0,0.6}
\definecolor{maudeOrange}{HTML}{c97c3a}
\definecolor{maudePurple}{HTML}{ad4db8}
\definecolor{maudeBlue}{HTML}{1e9694}
\definecolor{maudeGreen}{HTML}{90a834}
\lstdefinelanguage{Maude}{
    keywords={class, mod, endm, omod, op, endom, if, var, subsort, is, inc}, 
    escapeinside={(*@}{@*)},
    keywordstyle=\color{maudeOrange}\bfseries,
    comment=[l]{---}, 
    commentstyle=\color{gray}\itshape,
    morekeywords=[2]{red, reduce, result, crl, rl, eq, ceq}, 
    keywordstyle=[2]\color{maudePurple}\bfseries, 
    morekeywords=[3]{true, false}, 
    keywordstyle=[3]\color{maudeGreen}\bfseries, 
    sensitive=true, 
}

\newcommand{\eventually}{\lozenge\,}
\newcommand{\globally}{\square\,}

\begin{document}

\title{Towards the Coordination and Verification of Heterogeneous Systems with Data and Time}

\author{
\IEEEauthorblockN{Tim Kräuter \hypersetup{pdfborder={0 0 0}}\orcidlink{0000-0003-1795-0611}}
\IEEEauthorblockN{Adrian Rutle \hypersetup{pdfborder={0 0 0}}\orcidlink{0000-0002-4158-1644}}
\IEEEauthorblockN{Yngve Lamo \hypersetup{pdfborder={0 0 0}}\orcidlink{0000-0001-9196-1779}}
\IEEEauthorblockA{tkra@hvl.no, aru@hvl.no, yla@hvl.no\\
Western Norway University of Applied Sciences\\
Bergen, Norway}
\and
\IEEEauthorblockN{Harald König \hypersetup{pdfborder={0 0 0}}\orcidlink{0000-0001-6304-6311}}
\IEEEauthorblockA{harald.koenig@fhdw.de\\
University of Applied Sciences, FHDW\\
Hanover, Germany\\
Western Norway University of Applied Sciences\\
Bergen, Norway}
\and
\IEEEauthorblockN{Francisco Durán \hypersetup{pdfborder={0 0 0}}\orcidlink{0000-0001-5864-8094}}
\IEEEauthorblockA{fdm@uma.es\\
University of Málaga\\
Málaga, Spain}
}

\maketitle

\begin{abstract}
Modern software systems are often realized by coordinating multiple heterogeneous parts, each responsible for specific tasks.
These parts must work together seamlessly to satisfy the overall system requirements.
To verify such complex systems, we have developed a \textit{non-intrusive} coordination framework capable of performing \textit{formal analysis} of \textit{heterogeneous} parts that \textit{exchange data} and include \textit{real-time} capabilities.
The framework utilizes a linguistic extension---which is implemented as a central broker and a domain-specific language---for the integration of heterogeneous languages and coordination of parts.
Moreover, abstract rule templates are reified as language adapters for non-intrusive communications with the broker.  
The framework is implemented using rewriting logic (Maude), and its applicability is demonstrated by verifying certain correctness properties of a heterogeneous road-rail crossing system.
\end{abstract}



\section{Introduction}

Software systems are integral to nearly every aspect of modern life.
To meet their ever-growing requirements, these systems are often made by coordinating separate parts, which are implemented using the most appropriate tools for each of them.
Consequently, modern software systems consist of multiple heterogeneous parts, each responsible for specific tasks that must work together to fulfill the overall system requirements.
This rising complexity has made the development of software systems more challenging, while their ubiquitous nature amplifies their need for safety, reliability, and correctness~\cite{clarkeHandbookModelChecking2018}.

To verify properties---for example, safety and reliability---of systems consisting of \textit{heterogeneous} parts that \textit{exchange data} and can involve \textit{real-time} capabilities, we have developed a coordination framework designed for \textit{formal analysis}.
Here, heterogeneous means that each system part may be specified using a different \textit{real-time} \textit{behavioral} modeling language.
We refer to behavioral modeling languages as languages that specify the dynamic aspects of a system, such as statecharts, Petri Nets, and process models.
In contrast, structural modeling languages, such as UML class diagrams or entity-relationship models, focus on data representation.
Behavioral languages can include \textit{real-time} features, where actions are influenced by the passage of time.
This means one can define when an action can be executed and how long it takes to complete.
For instance, an action may occur periodically or after a specific time while in a particular state.
In addition, \textit{data exchange} is a unique feature currently missing in other coordination frameworks~\cite{krauterBehavioralConsistencyMultimodeling2023,varalarsenBCOolBehavioralCoordination2016,varalarsenBehavioralCoordinationOperator2015}.

Our coordination framework enables integration while upholding \textit{separation} of the different parts of the system by using a mediator, referred to as \textit{broker}.
Embracing separation makes the framework suitable for a wide range of scenarios and facilitates the addition of new modeling languages.

The main idea behind our approach is based on three key ingredients, namely \textit{language integration}, \textit{coordination}, and \textit{verification}, which are depicted in \autoref{fig: approach-overview} and explained in more detail in \autoref{sec: coordination-framework}:

\begin{figure*}[htb]
    \centering
    \includegraphics[width=.85\textwidth]{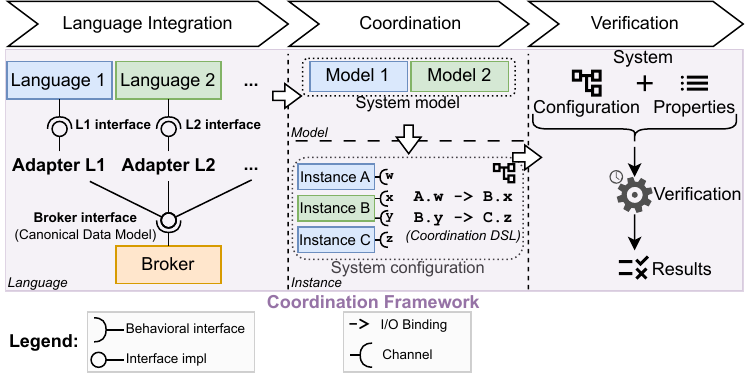}
    \caption{Overview of the Approach}
    \label{fig: approach-overview}
\end{figure*}

\noindent\textbf{Language Integration}.
Integrating a behavioral language into our framework requires implementing a \textit{language adapter}.
The adapter mediates between the behavioral language and the common language of the \textit{broker}.
Thus, we can use new languages by defining a \textit{behavioral interface} for the given language and then building an adapter that uses it to communicate with the broker.
The interface definition remains non-intrusive, leveraging a \textit{linguistic extension} for each modeling language.
Each adapter involves aligning the data model of the specific language with the broker's \textit{canonical data model} by defining data transformation functions.

\noindent\textbf{Coordination}. 
The system model consists of individual models conforming to the previously integrated behavioral languages.
These models are then instantiated and coordinated to create a \textit{system configuration}.
Since there might be multiple instances of a model, coordination is defined at the instance level.
Coordination relies on the \textit{channels} of each instance, i.e., specific entities where data can be read from or written to as identified in the behavioral interface of the corresponding language.
In our framework, we provide a domain-specific language (DSL) to define Input/Output (I/O) bindings, i.e., asynchronous data exchange between the channels.

\noindent\textbf{Verification}.
Once a system configuration is available, global properties of interest, i.e., \textit{system properties}, can be defined and checked.
In our framework, the verification takes \textit{real-time features} into account. 
In this work, we propose using reachability analysis and Linear Temporal Logic (LTL) model checking.
The result of the analysis is either a confirmation that the property holds or the identification of a counterexample.

The coordination framework provides an architecture built on general concepts like language adapters, I/O bindings, and a canonical data model.
To function, this framework must be instantiated with a formalism that can express these concepts and support verification involving data exchange and real-time features.
Once instantiated, it allows the analysis of heterogeneous systems by applying the methods of the chosen formalism. 
As explained in \autoref{sec: implementation-and-verifying-the-use-case}, our framework has been developed using rewriting logic and is implemented in Maude~\cite{manuelclavelAllMaudeHighPerformance2007}.
Maude allows us to provide semantics to different modeling languages and to formally check properties using its reachability analysis tool and model checker for real-time systems~\cite{ekerMaudeLTLModel2004,olveczkyRealTimeMaudeIts2014}. 
Our framework may also be implemented using other formalisms provided that the following requirements are satisfied:
\begin{enumerate}
    \item coping with the (operational) semantics of the modeling languages used to specify the different parts,
    \item supporting data exchange between the used modeling languages,
    \item modeling time and its passage affecting the dynamics of the global system, and
    \item providing analysis capabilities for the global system.
\end{enumerate}

The main contributions are summarized as follows.
\textbf{(i)} Our coordination framework provides a methodology to define coordination between multiple \textit{heterogeneous} behavioral models in a \textit{non-intrusive} manner, allowing for formal \textit{analysis} of the resulting \textit{global system}.
\textbf{(ii)} The framework enables \textit{data exchange} between different languages, a capability missing in previous coordination frameworks~\cite{krauterBehavioralConsistencyMultimodeling2023,varalarsenBCOolBehavioralCoordination2016,varalarsenBehavioralCoordinationOperator2015}. 
\textbf{(iii)} Moreover, the framework supports \textit{real-time functionality} while maintaining a clear separation between the integrated languages.

The remainder of the paper is structured as follows.
First, we provide the necessary background for our contribution in \autoref{sec: background}.
Then, we introduce a use case to motivate our framework in \autoref{sec: use-case}.
Afterward, we describe the coordination framework in \autoref{sec: coordination-framework} before instantiating it using rewriting logic (Maude) to obtain a concrete implementation, which we apply to the use case in \autoref{sec: implementation-and-verifying-the-use-case}.
Finally, we conclude in \autoref{sec: conclusion}.

\section{Background} \label{sec: background} 

Multiple research fields have investigated the coordination, verification, or simulation of software or even the combination of software and hardware systems.
In this section, we provide a brief overview of the three related research areas: \textit{coordination languages}, \textit{architecture description languages}, and \textit{co-simulation}.

\textbf{Coordination languages} can be broadly categorized into two families.
The Linda approach enables communication between software programs by providing a global shared memory, commonly called a tuple space, along with operations to read and write to this shared space~\cite{carrieroLindaContext1989}.
The coordination languages of the Linda approach are usually embedded in a general-purpose programming language~\cite{ciattoTwentyYearsCoordination2018,nixonTuplespacebasedComputingSemantic2008,rossiTuplebasedTechnologiesCoordination2001}.
In contrast to \textit{data-driven} coordination in the Linda approach, a family of \textit{control-driven} languages emerged~\cite{ciattoTwentyYearsCoordination2018} including for example Manifold~\cite{arbabOverviewManifoldIts1993,papadopoulosModellingActivitiesInformation1998} and REO~\cite{arbabReoChannelbasedCoordination2004}.
These languages define interfaces, i.e., ports for each entity, which can be connected to facilitate communication~\cite{papadopoulosCoordinationModelsLanguages1998}.

The goal of \textbf{Architecture Description Languages} (ADLs) is to describe the overall structure of a system by focusing on high-level system components and their connections~\cite{clementsSurveyArchitectureDescription1996,medvidovicClassificationComparisonFramework2000,medvidovicFrameworkClassifyingComparing1997}.
One uses an ADL to define which \textit{components} and \textit{connectors} exist before combining them in a concrete \textit{architectural configuration} to describe a system's structure~\cite{medvidovicClassificationComparisonFramework2000}.
Often, ADLs support verification, such as checking if an architectural configuration is free of deadlocks and starvation, by employing process algebras such as CCS, CSP, and $\pi$-calculus~\cite{ozkayaAreWeThere2013}.

\textbf{Co-Simulation} approaches facilitate the information exchange between multiple simulations running concurrently, ensuring temporal relationships.
A simulation is based on a simulation unit with black-box behavior but a predefined simulation interface of inputs and outputs.
Typically, an \textit{orchestrator} uses these interfaces to transfer data between simulations and dictate the passage of simulated time~\cite{gomesCoSimulationSurvey2019}.
Co-simulation approaches are classified into \textit{Discrete Event} (DE), \textit{Continuous Time} (CT), or \textit{Hybrid} when they combine both DE and CT elements.
At present, the two primary standards for co-simulation are the \textit{Functional Mock-up Interface} (FMI)~\cite{modelisarFunctionalMockupInterface2023} and the \textit{High Level Architecture} (HLA)~\cite{dahmannHighLevelArchitecture1997}.

The methodology in our approach outlines a \textbf{coordination framework}.
The unique characteristic of coordination frameworks is that they embrace \textit{model heterogeneity} by operating on the language level~\cite{varalarsenBehavioralCoordinationOperator2015}, i.e., not only providing one formalism or modeling language that must be used exclusively, such as coordination languages and ADLs.
ADLs typically verify only predefined properties and lack support for custom states and temporal logic formulas.
We do not consider co-simulation approaches as coordination frameworks since they compose executable programs (simulation units) that embrace heterogeneity at the execution level, not the model level.
Coordination frameworks described in the literature include Ptolemy~\cite{ptolemaeusSystemDesignModeling2014}, BCOoL~\cite{varalarsenBCOolBehavioralCoordination2016,varalarsenBehavioralCoordinationOperator2015}, and~\cite{krauterBehavioralConsistencyMultimodeling2023}.

However, because Ptolemy is focused on execution, it relies on a general-purpose programming language, which limits the ability to verify the coordinated system formally.
The other coordination frameworks are built on sound formalisms.
Still, they lack support for data exchange during coordination, which is essential and a natural form of communication in modern software systems.
Both approaches identify data exchange as a key focus for future work~\cite{krauterBehavioralConsistencyMultimodeling2023,varalarsenBCOolBehavioralCoordination2016,varalarsenBehavioralCoordinationOperator2015}.
Our coordination framework enables formal analysis of heterogeneous systems, incorporating \textit{data exchange} and \textit{real-time capabilities} while remaining \textit{non-intrusive} and upholding separation.

Our contribution focuses on software systems that can include real-time characteristics.
We exclude CT or hybrid systems, typically involving hardware components modeled by differential equations.
By applying the principles of Real-Time Maude, we can model continuous-time software systems that can be ``discretized" using time-sampling strategies available in Maude~\cite{olveczkyRealTimeMaudeIts2014}.

\section{Use Case} \label{sec: use-case}

In this section, we present a use case to illustrate the motivation behind our coordination framework.
We begin by introducing the use case, followed by explaining its specification.
Finally, we discuss several properties that must be verified for the specification.

\subsection{Description} \label{subsec: description}

The use case is a traffic management system for a \textit{level crossing} (also called road-rail crossing).
Level crossings account for over 400 accidents in the European Union every year~\cite{europeanunionagencyforrailways.ReportRailwaySafety2024}. 
The challenge with level crossings is that trains have a much larger mass relative to their braking capability and, thus, a far longer braking distance than road vehicles.
As a result, trains typically do not stop at a level crossing and depend on vehicles and pedestrians to clear the tracks beforehand.
\textit{Passive} level crossings are only equipped with a traffic sign and are associated with a higher number of accidents compared to \textit{active} crossings~\cite{europeanunionagencyforrailways.ReportRailwaySafety2024}.%

\begin{figure}
    \centering
    \includegraphics[trim={0 1.2cm 0 1.5cm},clip,width=0.489\textwidth]{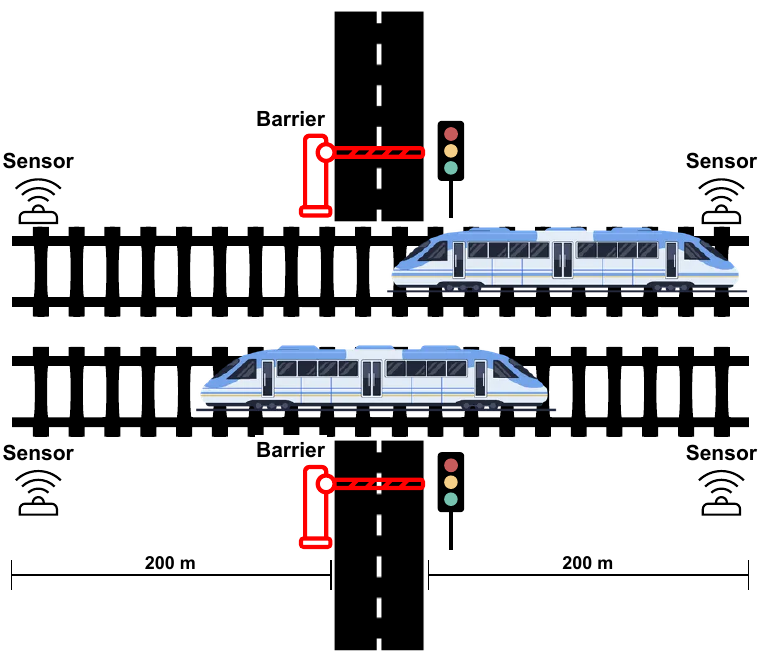}
    \caption{Use case: Active Level Crossing (two train tracks)}
    \label{fig: use-case-overview}
\end{figure}

This use case aims to design an \textit{active} level crossing system that warns cars about incoming trains using traffic lights and barriers.
\autoref{fig: use-case-overview} provides an overview of the scenario with two train tracks.
The case study was kept as simple as possible (trains do not communicate with the system) to illustrate the key concepts of our approach. 
In this case, the primary objective is to ensure that the proposed specification enables cars and trains to \textit{safely pass} through the crossing.

\subsection{Specification} \label{subsec: specification}

\begin{figure}[htb]
    \centering
    \includegraphics[width=0.45\textwidth]{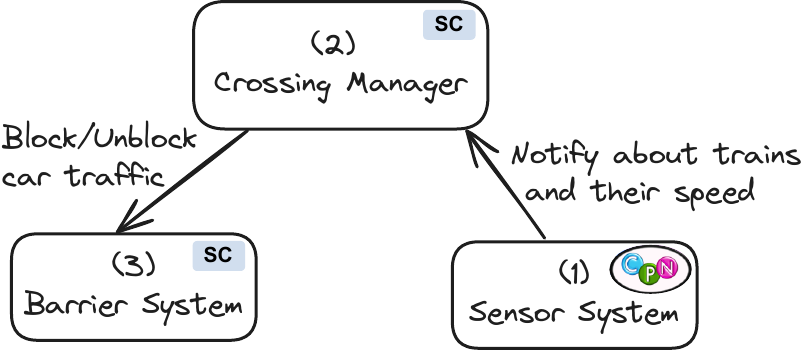}
    \caption{System parts in the use case}
    \label{fig: use-case-composition}
\end{figure}

The case study is realized by coordinating three system parts specified using two different modeling languages, namely Colored Petri Nets (CPN)~\cite{jensenColouredPetriNets2009} and statecharts.
This corresponds to the model level in the coordination step in \autoref{fig: approach-overview}.
We will explain the system parts as sketched in the diagram in \autoref{fig: use-case-composition}.
By default, all distances will be expressed in meters, time in seconds, and speed in meters per second.

\begin{figure}[htb]
    \centering
    \includegraphics[trim={0 0 0 0.2cm},clip,width=0.489\textwidth]{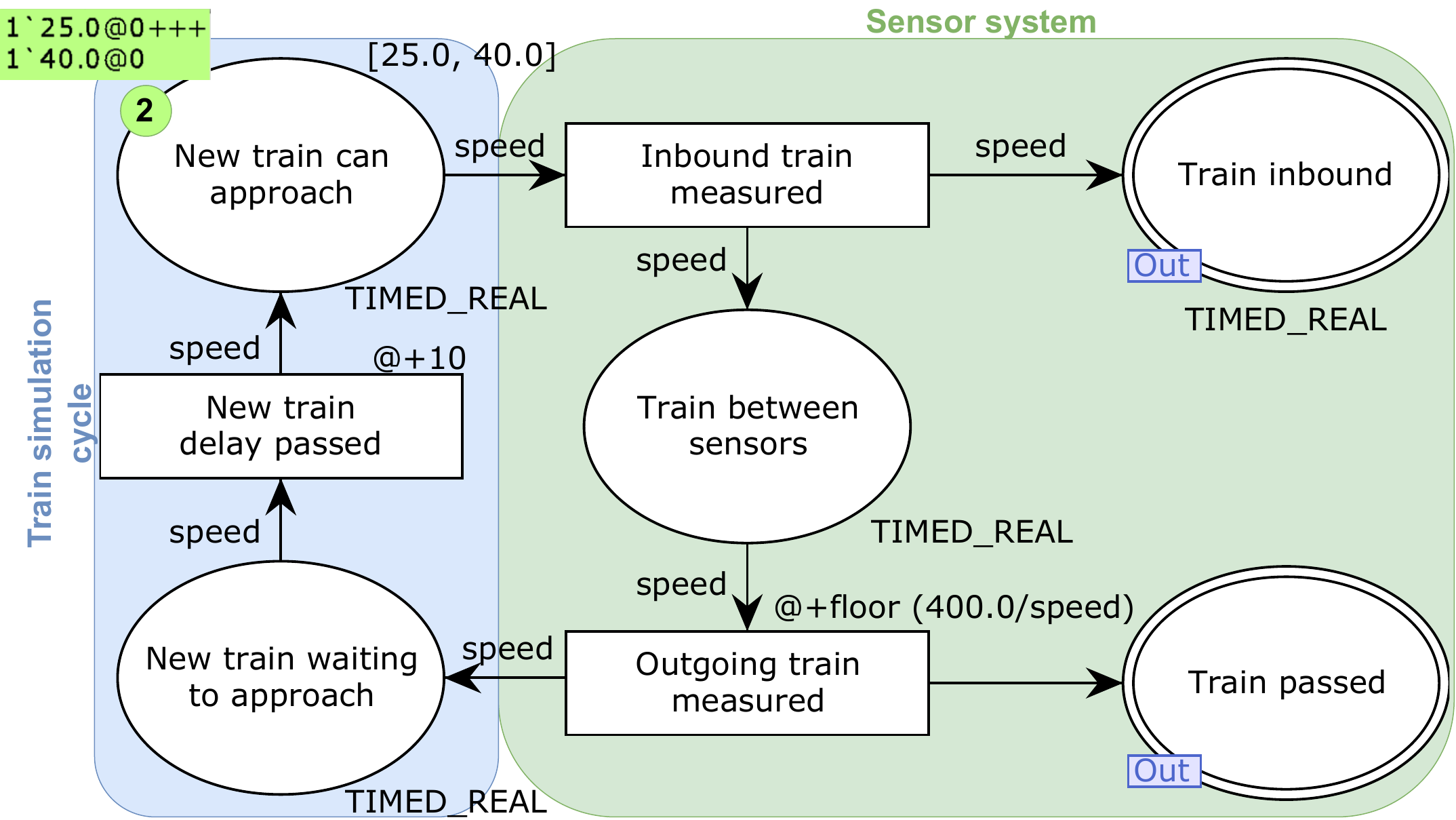}
    \caption{Train simulation \& Sensor system (CPN)}
    \label{fig: sensor-cpn}
\end{figure}

\paragraph{The sensor system} shown in \autoref{fig: use-case-overview} detects the speed of incoming and outgoing trains.
In general, trains can arrive in any order and at different speeds, and there may be level crossings with either one or multiple tracks.
To model train behavior flexibly, the model was developed using CPN as shown in \autoref{fig: sensor-cpn}.
It consists of two parts: a train simulation (blue) and the sensor system (green).

In a CPN, places are represented with ellipses and transitions with rectangles. 
Each token carries a data value that belongs to a given type, indicated at the bottom right of a place (\texttt{TIMED\_REAL} means real value and a timestamp).
In the state in \autoref{fig: sensor-cpn}, the place \texttt{New train can approach} (upper left corner) has two tokens with values 25 and 40 (train speed). 
Transitions can have a time inscription displayed at the top right, increasing the timestamp of produced tokens.
For example, when a token moves from \texttt{New train waiting to approach} to \texttt{New train can approach}, its timestamp increases by 10.
The arc labels are expressions that reference the values of tokens as they traverse the arcs. In the given use case, they ensure that the measured speed is transmitted correctly.
Finally, the places with double lines (the two on the right side of the figure) are \textit{port places}, representing the interface through which the model communicates with its surroundings.
The blue tag indicates the place's type; in this case, both are output ports.

The number of initial tokens in the CPN model determines the number of tracks being modeled, assuming that there is only one train at a constant speed on each track at any given time (25 m/s and 40 m/s in \autoref{fig: sensor-cpn}).
The two main transitions are \texttt{Inbound train measured} and \texttt{Outgoing train measured}, which refer to the sensors detecting a train as it enters or leaves the 400-meter corridor.
Sensors are assumed to be placed on the tracks 200 meters on each side of the barriers.
These transitions produce tokens in \texttt{Train inbound} and \texttt{Train passed}, which are the above-mentioned interfaces.
After a short cooldown period (see \texttt{New train delay passed}), new trains can start approaching the crossing, resulting in an endless cycle of trains passing over time; see the blue train simulation part on the left of \autoref{fig: sensor-cpn}.
Adjusting either the number of tokens or the train simulation within the CPN model can change the simulated behavior.

\begin{figure*}[htb]
    \centering
    \includegraphics[width=0.8\textwidth]{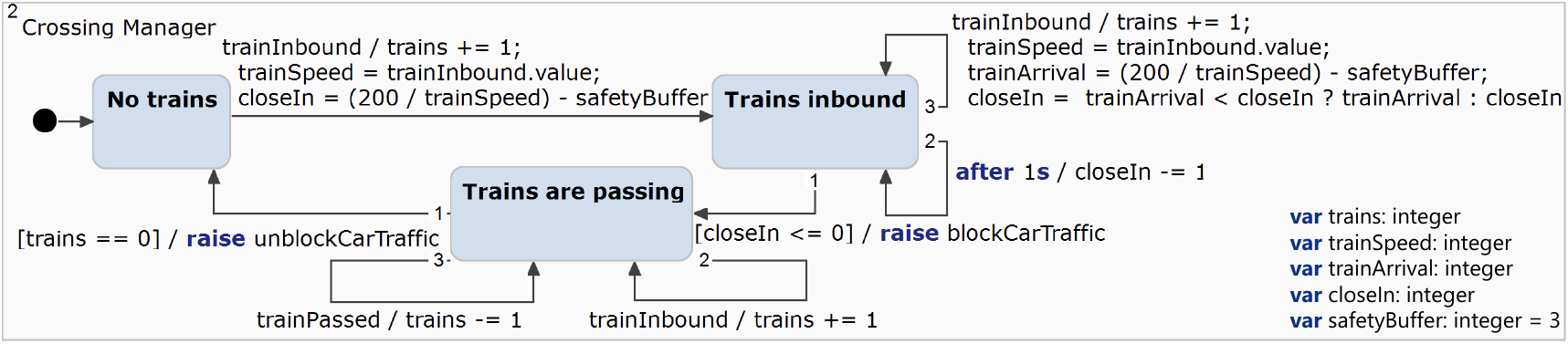}
    \caption{Crossing Manager (statechart)}
    \label{fig: crossing-manager}
\end{figure*}

\paragraph{The crossing manager} monitors the information from the sensors to detect whether any trains are within the 400-meter zone.
Based on this, it manages car traffic by sending signals to the barrier system.
The crossing manager is specified as a statechart in \autoref{fig: crossing-manager}.
Transitions are defined as usual using three optional components in the following format: \textit{trigger} \textit{[guard]} / \textit{effect}.
Here, the \textit{trigger} can be either time-based or event-driven, the \textit{guard} is a boolean expression, and the \textit{effect} consists of a series of statements (separated by semicolons) that raise an event or modify the statechart's variables.

The system starts in the state \texttt{No Trains} and transitions to the state \texttt{Trains inbound} when a \texttt{trainInbound} event is received. It also tracks the number of trains in the corridor using the \texttt{trains} counter. Additionally, it calculates how many seconds it will take for the trains to reach the level crossing, factoring in a safety margin: $\textit{closeIn} =200/\textit{trainSpeed} - \textit{safetyBuffer}$.

In the \texttt{Trains inbound} state, the \textit{closeIn} variable decreases every second until it reaches 0. At that point, the system transitions to the \texttt{Trains are passing} state, triggering the \texttt{blockCarTraffic} event.
However, additional trains may arrive while still in the \texttt{Trains inbound} state and could reach the level crossing before the previous train. In such cases, the \texttt{closeIn} variable must be updated accordingly.

In the \texttt{Trains are passing} state, the \texttt{trains} variable is either decreased when trains pass or increased when trains approach.
When the variable reaches 0, the \texttt{unblockCarTraffic} event is triggered, and the system transitions to the \texttt{No Trains} state, allowing the process to restart.
The events \texttt{blockCarTraffic} and \texttt{unblockCarTraffic} represent the data flowing from the crossing manager to the barrier system in \autoref{fig: use-case-composition}.

\begin{figure}[htb]
    \centering
    \includegraphics[width=0.489\textwidth]{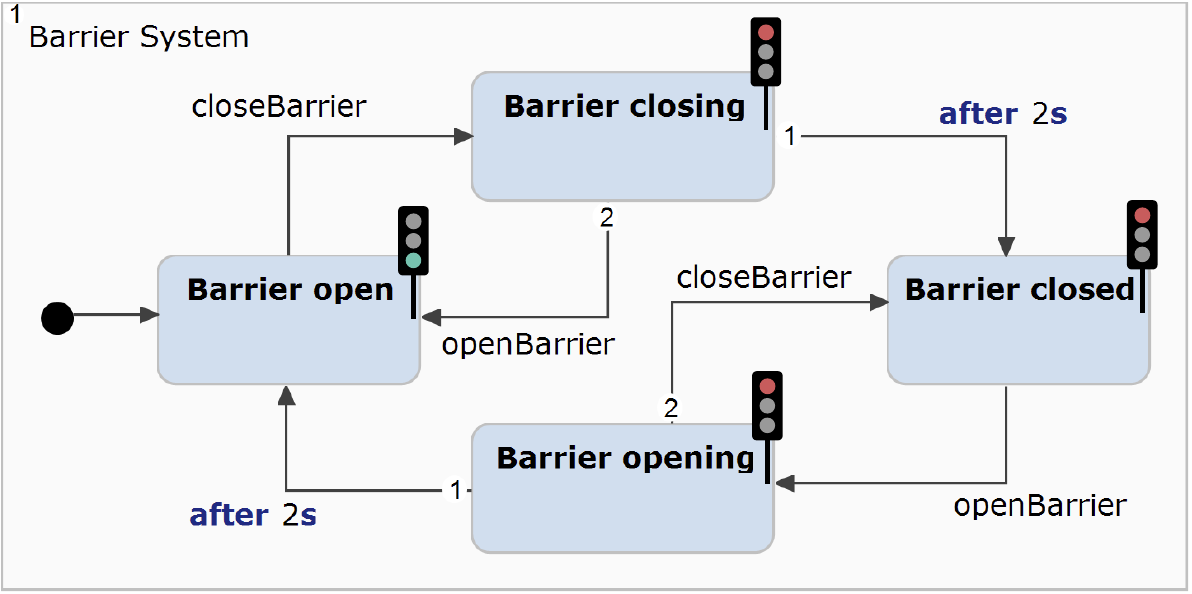}
    \caption{Barrier System (statechart)}
    \label{fig: barrier-system}
\end{figure}

\paragraph{The barrier system} manages car traffic by controlling the barriers and corresponding traffic lights.
The system is specified as a statechart in \autoref{fig: barrier-system}.
It depends on two \textit{external} events, \texttt{openBarrier} and \texttt{closeBarrier} (incoming data in \autoref{fig: use-case-composition}), which trigger an intermediate state lasting two seconds, representing the time required for the barrier to move.

All three systems---the sensor system, barrier system, and crossing manager---must work together to ensure safe and efficient traffic management at the level crossing.
This is more complex than it appears in \autoref{fig: use-case-composition}, as CPN models typically do not interact natively with statecharts. 
Moreover, even the event names in the statecharts do not align, such as \texttt{blockCarTraffic} and \texttt{closeBarrier}.
To address this challenge, the modeling languages must be integrated and proper coordination must be established, as illustrated in \autoref{fig: approach-overview}.
In the next section, we describe the intended verification to ensure correct coordination.

\subsection{Verification} \label{subsec: verification}

We aim to analyze the proposed specification from the previous section to identify potential issues early on.
To ensure the system behaves as intended, we focus on verifying \textit{two key properties}.
These properties are expressed using Linear Temporal Logic (LTL), employing state-based model checking as our primary verification method.
Nonetheless, in general, other analysis approaches may also be worth exploring.

First, we must ensure that the system does not create unsafe situations, such as allowing trains to pass while the barriers remain open.
This safety property can be expressed in LTL with Property~\eqref{eq:trafficSafety}, which expresses that a train never passes when the barriers are open.
\begin{equation}
    \globally \neg \,(\textit{Barriers-open} \, \land \textit{Train-passing}) \label{eq:trafficSafety} \\
\end{equation}

Second, we must ensure that when the barrier closes, it will eventually reopen so that car traffic is not indefinitely halted and Property \eqref{eq:trafficSafety} is trivially upheld.
This requirement is captured by Property~\eqref{eq:barriersReopen}, which represents a specific case of the \textit{response} pattern~\cite{dwyerPatternsPropertySpecifications1999}.
This pattern dictates that once the system enters the first state (\textit{Barriers-closed}), it must eventually transition to the second state (\textit{Barriers-open}).
\begin{equation}
    \globally \,(\textit{Barriers-closed} \, \to \eventually \textit{Barriers-open}) \label{eq:barriersReopen}
\end{equation}

Verifying these system properties is challenging for several reasons, and as a result, none of the approaches, even coordination frameworks discussed in \autoref{sec: background}, currently support this.
First, the system consists of three parts which are designed using \textit{heterogeneous modeling languages} (statecharts and CPN in our case study).
Second, \textit{data is exchanged} between these parts (train speed in our example), which affects how the system behaves (waiting times).
Third, the system's behavior is influenced by \textit{global time}, which needs to remain consistent across the different modeling languages.
Finally, we do \textit{not} want to be \textit{intrusive}, meaning imposing model changes or requiring a specific modeling language.

\section{Coordination Framework} \label{sec: coordination-framework}

The driving design goals for our coordination framework are \textit{non-intrusiveness} and upholding the \textit{separation} of participating models and instances.
Furthermore, we want the integration of additional languages to require minimal effort, for which separation across multiple key aspects is needed.
Separation, or loose coupling, is essential in software architecture because it simplifies making changes to the system by minimizing interdependencies~\cite{hohpeEnterpriseIntegrationPatterns2004,martinCleanArchitectureCraftsmans2017}. 

We focus on the separation of the following key aspects: \textit{languages}, \textit{models}, \textit{time}, and \textit{data}.
Firstly, each system part should be flexible in choosing the most suitable modeling language, provided the underlying formalism can support the language.
Secondly, every part of the system, i.e., each model, must be able to evolve independently.
Thirdly, each model should be able to use its native real-time features while time progresses uniformly during execution for all model instances, never skipping real-time events.
Finally, each model should be able to use its own data model yet still be able to exchange data with other system parts.
In the following sections, we detail our approach by explaining the key aspects highlighted in \autoref{fig: approach-overview} and explain how it meets the design goals of non-intrusiveness and separation.

Language integration, coordination, and verification are distinct tasks, each typically handled by different roles.
For instance, a language engineer is responsible for language integration, a system architect manages coordination, and the quality assurance team oversees the verification process.

\subsection{Language Integration}
\begin{figure*}[htb!]
    \centering
    \includegraphics[width=1\textwidth]{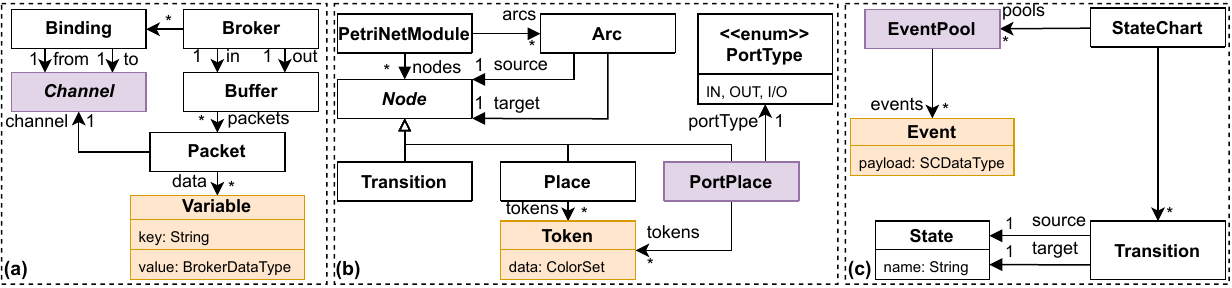}
    \caption{Broker Metamodel \textbf{(a)} and CPN/Statechart Metamodel excerpts \textbf{(b)}/\textbf{(c)}}
    \label{fig: broker-metamodels}
    \vspace{-6mm}
\end{figure*}

To facilitate coordination while upholding separation, we introduce a central \textit{broker}, which acts as a mediator.
We define the metamodel for the broker on the left in \autoref{fig: broker-metamodels}.
As shown in its metamodel, the broker is just a collection of bindings, which, as we will see below, are handled independently, without interactions between them.
To integrate heterogeneous languages, we use a \textit{linguistic extension}~\cite{atkinsonRearchitectingUMLInfrastructure2002,delaraGenericMetamodellingConcepts2010}.
The purple part of the broker metamodel defines the linguistic extension, which is used to augment behavioral languages non-intrusively.
It introduces the concept of a \texttt{Channel}, which refers to any location where data can either be read from or written to.
Linguistic extension is a widely used technique in the multi-level modeling domain, where new concepts can be added a posteriori to existing metamodels at any level of a metamodeling hierarchy.
In our case, we use the broker metamodel to define the concept (\texttt{Channel}), which can be used in the participating metamodels.
The channels are used to identify the behavioral interface for the participating languages, e.g., \texttt{PortPlace} for CPNs and \texttt{EventPool} for statecharts are both typed by \texttt{Channel} (see the purple colored elements in \autoref{fig: broker-metamodels}).

The broker metamodel also introduces \texttt{Packet} containing data in a \textit{canonical data model}~\cite{hohpeEnterpriseIntegrationPatterns2004} to enable data exchange among coordinated models while upholding separation.
The \textit{canonical data model} serves as a common intermediate to enable conversion between the varying data models of the used languages.
Data-related concepts are shown in orange in \autoref{fig: broker-metamodels}.
A packet is contained in an input or output \texttt{Buffer}, which holds the packet until it is transmitted to the recipient. 
A \texttt{Packet} contains \texttt{data}, i.e., \texttt{Variable}s, which are key-value pairs and references the \texttt{Channel} from which it was ingested or to which it is about to be delivered.

The broker has a set of bindings that connect \texttt{Channel}s.
As mentioned earlier, a \texttt{Channel} is a connection point in a given language that the broker can use to read or write.
If a binding is defined between two channels, the broker will facilitate coordination, i.e., data exchange.
The broker must know how to interface with each behavioral language to adapt between the heterogeneous models while maintaining language separation.
By augmenting a specific element in a metamodel with the concept of \texttt{Channel}, made possible by the linguistic extension, one can define the behavioral interface for that metamodel~\cite{varalarsenBehavioralCoordinationOperator2015}.

Language integration (see \autoref{fig: approach-overview}) can be further subdivided into two steps.
First, one must identify the behavioral interface for the new language, i.e., define which parts of the metamodel are exposed by extending \texttt{Channel}.
As a result, we obtain an \textit{augmented} metamodel, such as the metamodel for CPNs in \autoref{fig: broker-metamodels} (b), where \texttt{PortPlace} is identified as a \texttt{Channel} (purple coloring).
Since port places constitute the \textit{interface} through which a CPN exchanges tokens with its environments, we utilize port places as the behavioral interface for our framework.
More broadly, one can examine how different model instances within a language interact to determine its behavioral interface.
For example, in the case of statecharts, instances communicate using events dispatched through \textit{event pools}~\cite{objectmanagementgroupUnifiedModelingLanguage2017}.
Therefore, event pools serve as the behavioral interface for statecharts, i.e., an \texttt{EventPool} is a \texttt{Channel}, as highlighted by the purple color in the statechart metamodel in \autoref{fig: broker-metamodels} \textbf{(c)}.

Second, one utilizes the augmented metamodel to create a \textit{language adapter}, inspired by the adapter~\cite{gammaDesignPatternsElements1995} or channel adapter~\cite{hohpeEnterpriseIntegrationPatterns2004} patterns.
This language adapter facilitates the integration of heterogeneous languages by mediating between the specific language and the broker.

A key part of each language adapter is the translation between different data representations used by different languages.
To enable seamless data exchange, the broker employs a \textit{canonical data model} as a standardized format.
Each adapter is responsible for implementing two functions that handle the data translation process:
(i) The function $toBroker$ translates data from the specific data model to the broker's canonical data model, 
and (ii) the function $fromBroker$ translates data from the broker's canonical data model back to the specific data model.
These two functions should be inverse functions of one another.
Using a \textit{canonical data model} addresses the problem where normally the number of translators needed to enable communication between each pair of participants increases quadratically with the number of participants~\cite{hohpeEnterpriseIntegrationPatterns2004}.
The canonical data model, i.e., packets consisting of key-value variable pairs, is kept abstract but sufficient to showcase the framework's data exchange capabilities.

In addition to \textit{syntactic mapping} between different data models, as discussed so far, schema matching, i.e., \textit{semantic mapping}~\cite{cheathamSemanticDataIntegration2017}, is also vital to enable system interoperability.
Semantic mapping is needed since systems can represent the same information structurally differently.
One system might use different field names or even split information into two fields compared to another system.
Semantic mapping allows reconciling these differences when data is transferred between the systems.
In our coordination framework, semantic mapping can be included in the translation functions and customized for each binding or even generated from relations between the data models, often described graphically in data integration tools.
However, we do not investigate this further in this paper.

\begin{figure}[htb]
    \centering
    \includegraphics[width=0.241\textwidth]{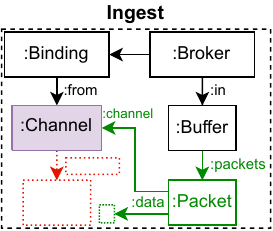}
    \includegraphics[width=0.241\textwidth]{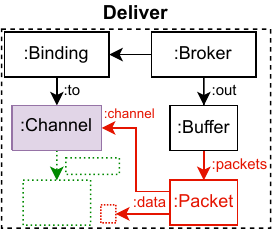}
    \caption{Language Adapter Rule Templates \textbf{Ingest} and \textbf{Deliver}}
    \label{fig: Template-LA-rules}
\end{figure}

Language adapters can be defined as follows.
A language adapter reads from the channels of model instances to create an intermediate packet, provided that a suitable binding source is defined (\textbf{ingest}).
Additionally, for binding targets, an adapter writes to the channels and removes the previously created packets (\textbf{deliver}).
These two actions can be described by rule templates~\cite{maciasMultilevelCoupledModel2019,sanchezcuadradoGenericModelTransformations2011}, using a Henshin-like notation~\cite{struberHenshinUsabilityFocusedFramework2017}, as shown in \autoref{fig: Template-LA-rules}, which can then be specialized for different languages to build a concrete language adapter.
The rules can be interpreted as follows: the black elements represent objects that will remain unchanged during the transformation.
The red elements will be deleted during the transformation process, while the green elements will be added.
The dotted elements need to be filled in for a specific language adapter.

\begin{figure*}[htb]
    \centering
    \includegraphics[trim={0 0.1cm 0 0},clip,width=0.9\textwidth]{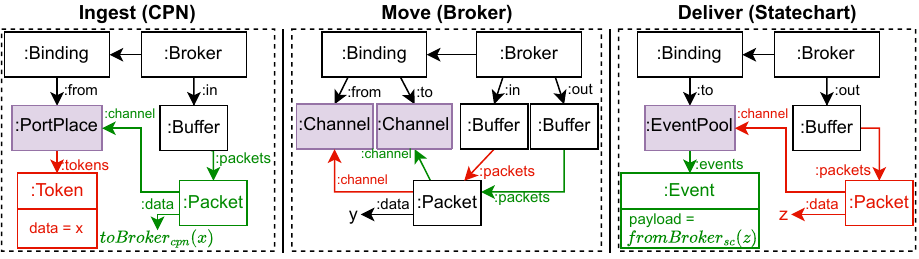}
        \caption{Broker and Language Adapter Rules}
    \label{fig: example-rules}
\end{figure*}

For instance, \autoref{fig: example-rules} shows adapter rules for CPN and statecharts, as well as the generic move rule for the broker.
The \textbf{ingest} rule for CPN is displayed on the left, and the \textbf{deliver} rule for statecharts is displayed on the right.
Both \textbf{ingest} and \textbf{deliver} rules require converting between different data models, for which we employ the previously defined translation functions.
Specifically, the functions $toBroker_{cpn}$ and $fromBroker_{sc}$ are used for translation.
Consequently, an event generated from a CPN token containing data $x$ will have the payload $fromBroker_{sc}(toBroker_{cpn}(x))$.

The \textbf{move} rule, depicted in the middle, demonstrates how the broker transfers packets from its input buffer to its output buffer, effectively isolating the language adapters.
Following the rules from left to right shows how a token is converted into an event, including associated data.

\textbf{In summary,} introducing a broker as a mediator achieves the \textit{separation} of languages, models, and data.
Due to the concept of language adapters and the canonical data model, we keep languages, models, and data as separated as possible.
Additionally, identifying a behavioral interface for each language through a \textit{linguistic extension} ensures \textit{non-intrusiveness}, as models do not have to be changed for coordination.
Notably, incorporating a new language into the framework involves a \textbf{one-time integration effort}.
We will now detail how time separation is achieved, and we guarantee consistent passage of time across heterogeneous formalisms.

\noindent\textbf{Time.}
Many modeling languages include the real-time functionality to specify, for example, that actions take a certain amount of time, are periodic, or can only happen after a specific amount of time.
Each language will provide its own real-time elements, and the behavioral semantics of these elements will be given as part of the semantics of the corresponding languages. 
For example, time inscriptions for CPNs~\cite{jensenColouredPetriNets2009} and real-time triggers in statecharts are employed in the use case.
However, to guarantee consistent time treatment, individual model instances that are part of the coordinated system cannot perform timed actions as they please since all actions must be consistent with time elapse.

Individual model instances cannot progress time independently.
There must be a model of time that is followed by all language definitions participating in the framework. 
As we will discuss in \autoref{sec: implementation-and-verifying-the-use-case}, our framework has been implemented using rewriting logic (concretely the Maude system). 
Therefore, we expect that the real-time features of the different language definitions adhere to the principles of Real-Time Maude~\cite{olveczkyRealTimeMaudeIts2014}.
We assume a \textit{global clock}, which will advance and then synchronize with the clocks of the individual instances.
The global clock is only changed by the \textit{tick} rule, meaning all other rules can be seen as \textit{instantaneous}. 
For instance, to model some duration, we will have a start action and an end action, both instantaneous. 
All time-related actions will have a timer or a scheduled time that will be used to fire them. 
Instead of keeping a centralized scheduled-time sorted list of actions, we will assume functions calculating the time to the first action, or the maximum amount of time that may elapse without an action happening (\textit{mte}), and another one applying the pass of time (\textit{delta)}. 
Given these functions, the passage of time can then be modeled by a unique tick rule that calculates the maximum time elapse without an executable action (\textit{mte}) and applies the \textit{delta} function to the entire system. 
Under these assumptions, the control of time can be easily split between the different parts of the system. 
The \textit{mte} function will be defined as the minimum of its results for each part of the system. 
Then, the delta function on the global system will simply consist of applying it to each of its parts. 
The definitions of the \textit{mte} and \textit{delta} functions for each modeling language will be part of the definition of each language adapter.
Time-related actions will then be fired when a timer reaches time zero or a clock matches a local or global clock.
As already said, this approach to real-time is based on Real-Time Maude~\cite{olveczkyRealTimeMaudeIts2014}, which is similar to the models used for DE-based co-simulation methods~\cite{gomesCoSimulationSurvey2019}, where an orchestrator controls global time.

\subsection{Coordination}

The second step in our approach (see \autoref{fig: approach-overview}) is the \textit{coordination} of the system parts (models).
This step can be further subdivided into the following three steps.
First, the \textit{system model} is created, which consists of individual models conforming to the previously integrated behavioral modeling languages.

Second, one instantiates the system model by instantiating its respective individual models.
In the use case, each model (the barrier system, sensor system, and crossing manager) has only one instance, although multiple instances of the same model are possible.

The third step is based upon the augmented metamodel obtained from the language integration step (see \autoref{fig: approach-overview}).
The linguistic extension defines each language's behavioral interface, specifying which channels can be linked via bindings.
We utilize a textual DSL to connect channels from different modeling languages uniformly.
For example, for the use case, we define the bindings as shown in \autoref{lst:bindings}.
Each line defines one binding, following the structure \textit{bindingSource} \textcolor{cyan}{\textbf{\texttt{->}}} \textit{bindingTarget} in our DSL.
Finally, the model instances, along with their corresponding bindings, constitute the \textit{system configuration}, which is utilized for verification.

\begin{lstlisting}[
label=lst:bindings,numbers=none,captionpos=b,
float=tb,
basicstyle=\ttfamily\midsize,
caption=Bindings for the use case]
Sensor.Train_inbound (*@  \textbf{\textcolor{cyan}{->}}  @*) Crossing.trainInbound,
Sensor.Train_passed (*@  \textbf{\textcolor{cyan}{->}}  @*) Crossing.trainPassed,
Crossing.unblockCarTraffic (*@  \textbf{\textcolor{cyan}{->}}  @*) Barrier.openBarrier,
Crossing.blockCarTraffic (*@  \textbf{\textcolor{cyan}{->}}  @*) Barrier.closeBarrier
\end{lstlisting}

\subsection{Verification}

\textit{Verification} in our approach (see (iii) in \autoref{fig: approach-overview}) is further subdivided into specification and verification of system properties.

First, one specifies the \textit{system properties} that must be verified based on the \textit{system configuration}.
To specify the properties of the system's global state, it is necessary to first define the \textit{state structure} for each participating modeling language, as explained, for example, in~\cite{krauterBehavioralConsistencyMultimodeling2023,krauterHigherorderTransformationApproach2024}.
The state structure, which is also needed for execution, leads to atomic propositions for each language that can be combined to express temporal logic properties for the global system, e.g., the above formulas \eqref{eq:trafficSafety} and \eqref{eq:barriersReopen}.
The global state is a tuple of the local states of each model based on the system configuration.
For example, the initial state in the use case is a triple consisting of the statechart start states and the two tokens in the place \texttt{Train prepared to approach}.

Second, we use the system properties and the system configuration from the previous steps to perform \textit{automatic} verification, for example, LTL model checking~\cite{clarkeHandbookModelChecking2018}, where the states are the tuples mentioned in the previous step.
This generates verification results, including counterexamples, if any properties are unsatisfied.

\section{Implementing \& Verifying the Use Case} \label{sec: implementation-and-verifying-the-use-case}

We implemented the framework using rewriting logic and its implementation in Maude~\cite{manuelclavelAllMaudeHighPerformance2007} as the foundational formalism.
The full implementation in~\cite{timkrauterArtifactsMODELS20252025} includes the use case in addition to definitions of Labeled Transition Systems (LTS)~\cite{clarkeHandbookModelChecking2018} and Business Process Model and Notation (BPMN)~\cite{objectmanagementgroupBusinessProcessModel2013}---in these cases without time or data---as well as several examples using different combinations of the available languages.
Maude meets the formalism requirements outlined in the introduction: it can handle a wide variety of operational semantics (1)~\cite{marti-olietRewritingLogicLogical1996,meseguerRewritingLogicSemantics2013}, supports the modeling of data and data exchange (2)~\cite{manuelclavelAllMaudeHighPerformance2007}, enables time modeling through Real-Time Maude~\cite{olveczkyRealTimeMaudeIts2014} (3), and offers built-in verification capabilities (4).
Verification in Maude includes techniques such as reachability analysis, explicit-state LTL model checking~\cite{ekerMaudeLTLModel2004}, and time-bounded LTL model checking~\cite{olveczkyRealTimeMaudeIts2014}.
We now present key implementation details from each step of our approach (see \autoref{fig: approach-overview}), using the use case as a practical example.

In Maude, object-oriented systems may be specified using the usual elements in object-oriented languages, which allows us to specify the models in \autoref{fig: broker-metamodels}. 
For instance, there is a class Broker with \texttt{in} and \texttt{out} attributes of type \texttt{Packets}, representing buffers of packets, and an attribute \texttt{bindings} of channel bindings.

\begin{lstlisting}[language=Maude,basicstyle=\ttfamily\midsize,numbers=none]
class Broker | in : Packets, 
               out : Packets, 
               bindings : Set{Binding} .
\end{lstlisting}
    
\subsection{Language Integration}

Each formalism integrated into the framework will have its own representation. 
For example, following the descriptions in \autoref{fig: broker-metamodels}, CPNs will be represented by classes \texttt{CPN} and \texttt{CPNInstance}. 
An object of class \texttt{CPN} has a set of places, a set of transitions, and a set of arcs. 
In some CPN objects, a subset of the places will be port places. 
An object of \texttt{CPNInstance} is associated with its CPN model (\texttt{cpn}) and has a marking (multiset of tokens).

\begin{lstlisting}[language=Maude,basicstyle=\ttfamily\midsize,numbers=none]
class CPN | places : Set{CPNPlace}, 
            transitions : Set{CPNTransition}, 
            arcs : Set{CPNArc} .
class CPNInstance | cpn : Oid, 
                    marking : CPNTokens .
\end{lstlisting}

Given appropriate class declarations, objects are then represented with syntax \mbox{\texttt{< O : C | Atts >}}, with \texttt{O} an object identifier, \texttt{C} a class identifier, and \texttt{Atts} a comma-separated set of attribute-value pairs with the form \texttt{a : v}, with \texttt{a} the attribute's name and \texttt{v} its value. 

The behavior of these objects is then specified by rewriting rules.
For the language integration, the specification includes rules \textbf{ingest} and \textbf{deliver} similar to those in~\autoref{fig: example-rules}. 
For example, the language adapter rule \textbf{ingest} for CPNs is implemented as shown in \autoref{lst:ingest}. 
Objects of classes \texttt{CPNId} of class \texttt{CPN}, \texttt{IId} of class \texttt{CPNInstance}, and \texttt{Br} of class \texttt{Broker} are involved in the rules. 
In it, we can see how the \texttt{CPNInstance} object \texttt{IId} corresponds to the \texttt{CPN} object \texttt{CPNId} (\texttt{cpn} attribute).
We can also see how, if there is a token in the place \texttt{PlaceId} in the marking of the CPN instance object, with a binding for \texttt{PlaceId}, the token is removed from the CPN instance and added in the \texttt{in} buffer of the broker object.
Notably, the data transformation to the canonical data model occurs in the operation \texttt{cpnToBroker} (lines 11 and 17) and is not fully shown in \autoref{lst:ingest}.
Furthermore, the rule presented in \autoref{lst:ingest} is a simplified version and does not account for the token's availability at the current global time.

\begin{lstlisting}[
language=Maude,
captionpos=b,
float=htb,
basicstyle=\ttfamily\midsize,
label=lst:ingest,
caption=CPN Language Adapter Rule \textbf{Ingest} in Maude~\cite{timkrauterArtifactsMODELS20252025}]
omod CPN-ADAPTER is 
 inc BROKER . 
 inc CPN-SEM .
 op cpnToBroker : CPNData -> VariableSet .
 ...
 rl [token_to_packet] :
    < CPNId : CPN | 
      places :
        (place(PlaceId, InArcs, OutArcs, Type),
         Places) >
    < IId : CPNInstance | cpn : CPNId, 
      marking : (token(PlaceId, Data, 0), Marking) >
    < Br : Broker | in : Packets, 
      bindings : (PlaceId -> ChannelId, Bindings) >
 (*@\textbf{\textcolor{maudeBlue}{=>}}@*) < CPNId : CPN | >
    < IId : CPNInstance | marking : Marking >
    < Br : Broker | 
      in : (packet(PlaceId, cpnToBroker(Data)),
            Packets) > .
endom
\end{lstlisting}

A similar Maude rule implements the \textbf{deliver} rule, specified in~\autoref{fig: example-rules}.
It is important to highlight that each adapter is implemented in its own Maude module, demonstrating how the separation of the languages is reflected at the code level.

To apply our implementation to the use case, we developed a language adapter for both CPNs and statecharts in~\cite{timkrauterArtifactsMODELS20252025}.
These language adapters are built on our broker metamodel (see \autoref{fig: broker-metamodels} (a)), and the identification of behavioral interfaces within the CPN and statechart metamodels (see \autoref{fig: broker-metamodels} (b)/(c)).
Implementing language adapters involves converting the data models of CPN and statecharts into the broker’s canonical data model.
Additionally, language integration ensures that the semantics of both CPN and statecharts are aligned with the externally provided global clock.

\noindent\textbf{Global Time Elapse.}
As already discussed in \autoref{sec: coordination-framework}, the model of time is provided by Real-Time Maude~\cite{olveczkyRealTimeMaudeIts2014}.
In summary, time passing is realized by a single \textit{tick rule}, which models the time elapse of the global clock based on the \textit{mte} (maximum time elapse) function,
and the \textit{delta} function, which applies the time pass to each model instance.
The key element in this model of time and the approach followed is that, as already said, the tick rule is the only rule that models the passage of time.
All other rules in the specification are instantaneous, meaning they can use time values but not advance time. 
This approach allows us to keep as many clocks and timers as necessary but with one single global clock that is used to synchronize the others. 
Moreover, the specification of the different formalisms gets simplified because all we have to do, regarding time, is to provide the appropriate equations defining the behavior of the \textit{mte} and \textit{delta} functions on its instances and time-related features. 

\subsection{Coordination}

The coordination is straightforward in Maude.
We define models and their instances separately, as illustrated in \autoref{lst:ingest}, where an instance links back to its model (\texttt{cpn} attribute).
We define each model from the use case, instantiate it once, and add bindings to obtain the system configuration. 
To add bindings, we use the broker metamodel from the language integration, which implements the bindings DSL we introduced earlier (see \autoref{lst:bindings}).

The Maude rules corresponding to the rules in \autoref{fig: example-rules} handle the coordination. 
The adapter definition identifies the elements used in the specific language for communication and how to handle them. 
In the case of the CPN, we have seen how the \textbf{ingest} rule is in charge of taking a token from a port place (a place for which there is a binding), and placing it into the input buffer of the broker. 
The \textbf{move} rule then moves packets from the input buffer to the output buffer. 
Finally, the adapter of the formalism on which the target part is specified will handle such a packet.
That is, it will take the packet and transform it into the communication element of the specified channel. 
For example, in our example, the target is specified as a statechart. 
Thus, the statechart adapter deletes the packet and creates an event in the target channel of the statechart.

\subsection{Verification}

In the case of Maude, we can perform verification using different tools, including reachability analysis, model checking and statistical model checking. 
Given that the system has an infinite state space, we need to use time-bounded LTL model checking~\cite{olveczkyRealTimeMaudeIts2014} to verify its behavior.
Although we could also check timed CTL properties, we chose time-bounded LTL for its simplicity and easier understanding.
To specify LTL properties, we first must define the necessary atomic propositions. 
For example, consider the property \eqref{eq:trafficSafety} specifying that a train never passes the crossing while the barriers are open. 
To specify this property, we can define the propositions \texttt{Train-passing} and \texttt{Barriers-open}.
As shown in \autoref{lst:verification}, given a satisfaction predefined operator \verb~_|=_~, we define a proposition by specifying when a state satisfies that proposition. 
For example, the \texttt{Train-passing} proposition is satisfied if there is a token in the \texttt{"Train passed"} place of the CPN instance object (lines 5-7). 
Similarly, the \texttt{Barriers-open} proposition is satisfied if the statechart is in the state \texttt{"Barrier open"} (lines 11-13). 

\begin{lstlisting}[
language=Maude,
numbers=left,
float=htb,
captionpos=b,
basicstyle=\ttfamily\midsize,
label=lst:verification,
caption=Atomic Propositions in Maude~\cite{timkrauterArtifactsMODELS20252025}]
mod PREDICATES is 
  inc BROKER-EX . 
  pr SATISFACTION .
  --- Sensor system propositions
  op Train-passing : -> Prop .
  ceq { < Id : CPNInstance | 
        marking :
            (token("Train passed", data(Int), T)) >
        C, GT } (*@\textbf{\textcolor{maudeBlue}{|=}}@*) Train-passing = true 
  if GT ge T .
  --- Barrier system propositions
  op Barriers-open : -> Prop .
  eq { < Id : SCInstance |
       state : scToken("Barrier open", T) >
       C, GT } (*@\textbf{\textcolor{maudeBlue}{|=}}@*) Barriers-open = true .
  --- Otherwise, propositions are false
  var S : System .
  var P : Prop .
  eq S (*@\textbf{\textcolor{maudeBlue}{|=}}@*) P = false [owise] .
endm
\end{lstlisting}

Given the propositions in \autoref{lst:verification}, we can then verify the desired Property \eqref{eq:trafficSafety} with the following command. 
\begin{lstlisting}[language=Maude,basicstyle=\ttfamily\midsize,numbers=none]
red modelCheck(system,
              []~(Barriers-open /\ Train-passing)) .
result Bool: true
\end{lstlisting}
\noindent Property \eqref{eq:barriersReopen} can be encoded similarly in Maude by defining the missing proposition \texttt{Barriers-closed} accordingly.
Both properties are fulfilled.

To use the model checker, in addition to the executability of the specification (i.e., the equational part of the specification must be terminating and Church-Rosser, and equations and rules must be coherent), the reachable state space must be finite. 
In this case, the search space is finite only if we limit the global time. 
The upper time bound for the verification is defined in the Maude rule that advances the global time in the system; it is not part of the property specification.

Atomic propositions can be defined in different Maude modules and then later combined into a system property as shown \autoref{lst:verification}.
By instantiating single models, adjusting bindings, or stubbing models, one can verify not only system properties but also properties for each individual model or a certain subset of the system model.

If a property is not fulfilled, the Maude model checker provides a counterexample, i.e., a sequence of states (set of objects) connected by transitions (rewriting steps).
These steps in the sequence reflect steps in terms of the operational semantics of the corresponding formalisms.
Our formal implementation strives to minimize the representation distance, ensuring near-zero cost when translating between a formalism's semantics and its Maude encoding.
Thus, going from a sequence of states back into a trace in the corresponding formalism is straightforward.
In the future, we plan to trace problems found by the model checker to the individual models used in our approach or the specified coordination, which might be faulty.

\section{Conclusion \& Future Work} \label{sec: conclusion}

We introduced a \textit{non-intrusive} coordination framework that can coordinate system parts described in \textit{heterogeneous} behavioral modeling languages into a \textit{global system} and allows for formal analysis of the system's properties.
The framework uses language adapters and a broker to integrate the heterogeneous languages, allowing for \textit{data exchange} and \textit{real-time capabilities} while upholding the \textit{separation} of the languages and models.
Furthermore, we implemented our framework using rewriting logic in Maude and used it to verify the correctness of an active level crossing system modeled with Colored Petri Nets and statecharts.
Additionally, the framework provides a general methodology for coordination, making it independent of any particular formalism for implementation; 
that is, the coordination DSL will ultimately hide the details of the underlying implementations.
It can be applied to software systems with real-time capabilities, i.e., continuous time that can be approximated or discretized by time-sampling in Maude.

As future research directions, we plan to improve our approach to address the following observations. 
The coordination DSL supports only binding instance channels; with user-friendliness in mind, it should also include syntax and constructs to support the specification of properties which the system configuration should satisfy---while hiding the underlying implementation.
Currently, our implementation in Maude is only applied to one specific use case, and further enhancements are needed to support all possible CPN and statechart features.
Nevertheless, it effectively demonstrates the overall architecture proposed in our approach and the proposed patterns for integrating various behavioral languages into our framework.
Following the same architecture, we investigated additional behavioral languages (BPMN, LTS) and synchronous communication besides asynchronous communication (see~\cite{timkrauterArtifactsMODELS20252025}).
Furthermore, the broker must not be a single, centralized component as it is currently presented, since multiple message brokers exist in a realistic distributed system. 
In addition, we want to make bindings multi-ary to model advanced communication patterns directly, such as broadcasting and publish/subscribe.
Currently, such patterns must be implemented in a dedicated model that duplicates messages together with multiple corresponding bindings.
Finally, we aim to validate our approach in a real scenario, not only the presented use case.

\bibliographystyle{IEEEtran}
\bibliography{bib}

\end{document}